\PassOptionsToPackage{x11names}{xcolor}
\documentclass[11pt,a4paper]{article}
\usepackage{jheppub}

\usepackage[normalem]{ulem}
\usepackage{amstext}
\usepackage{amssymb}
\usepackage{amsmath}
\usepackage{xcolor}
\usepackage{graphicx}
\graphicspath{{plots/}}
\usepackage{url}
\usepackage{color}
\usepackage{ulem}
\usepackage[utf8]{inputenc}
\pdfoutput=1
\usepackage{textcomp}
\usepackage{booktabs}
\usepackage{booktabs,siunitx,array,threeparttable}
\usepackage[T1]{fontenc}
\usepackage{ae,aecompl}
\usepackage{appendix}
\usepackage{slashed}
\usepackage{epsfig,amsfonts,mathrsfs,amsmath,amssymb,graphicx,color,slashed,multirow}
\usepackage{amsmath,latexsym,amssymb,graphicx,slashed,color,enumerate,url,cancel,gensymb}
\usepackage{textcomp}
\usepackage{multirow}

\usepackage[x11names]{xcolor}
\usepackage{orcidlink}
\usepackage{enumerate}

\usepackage{tikz-feynman}
\tikzfeynmanset{compat=1.1.0}
\usetikzlibrary{patterns,decorations.pathreplacing}
\usetikzlibrary{decorations.pathmorphing}
\usetikzlibrary{decorations.markings}
\usetikzlibrary{arrows,shapes}
\usetikzlibrary{shapes.misc}
\tikzset{
  gluon/.style={decorate, draw=black,
    decoration={coil,amplitude=4pt, segment length=4pt,aspect=0.7}} 
}
\tikzset{
  photon/.style={decorate, decoration={snake}},
}

\usepackage{textgreek} 
\usepackage{booktabs} 

\title{
First bounds on effective muon interactions \\ using the NA64$\mu$ experiment at CERN}

\author[a]{Paolo Crivelli~\orcidlink{0000-0001-5430-9394},} 
\author[b]{Josu Hernández-García~\orcidlink{0000-0003-0734-0879},}
\author[b]{Jacobo López-Pavón~\orcidlink{0000-0002-9554-5075},}
\author[b]{Víctor Martín Lozano~\orcidlink{0000-0002-9601-0347}, }
\author[b]{and Laura Molina Bueno~\orcidlink{0000-0001-9720-9764}}

\affiliation[a]{ETH Z\"{u}rich, Institute for Particle Physics and Astrophysics, CH-8093 Z\"{u}rich, Switzerland}
\affiliation[b]{Instituto de Física Corpuscular (IFIC), CSIC‐Universitat de València, Spain}

\emailAdd{paolo.crivelli@cern.ch} 
\emailAdd{josu.hernandez@ific.uv.es}
\emailAdd{jacobo.lopez@ific.uv.es}
\emailAdd{victor.lozano@ific.uv.es}
\emailAdd{laura.molina.bueno@cern.ch}

\abstract{We analyze how NA64$\mu$ can contribute to the global SMEFT program demonstrating that it can probe two effective four-lepton operators completely unbounded so far and break one of the current flat directions. Furthermore, we also study an extension of SMEFT that includes fermion singlets of the SM gauge group in the low energy field content. This effective field theory, usually dubbed $\nu$SMEFT, is well motivated by the observation of light neutrino masses and leptonic mixing. We find that NA64$\mu$ can constrain three unbounded four-fermion operators of the $\nu$SMEFT. We derive the current bounds on these operators and compute the future sensitivity.}

\arxivnumber{2511.11801}


\begin{document}

\maketitle

\section{Introduction}

Despite the intensive searches for New Physics at the high energy frontier, no sign of it has been found, therefore hinting that it might lie above the current reach of accelerators. Nevertheless, the effects of heavy new degrees of freedom can appear at low energies through higher-dimensional effective operators, parametrized by Effective Field Theories (EFTs). Among these, the Standard Model Effective Field Theory (SMEFT)~\cite{Buchmuller:1985jz,Grzadkowski:2010es} has become the canonical framework for probing indirect signs of heavy New Physics, while its extension to include light fermionic singlets, denoted as $\nu$SMEFT~\cite{Graesser:2007yj,Graesser:2007pc,delAguila:2008ir,Liao:2016qyd}, opens the door to studying scenarios involving dark sectors and Heavy Neutral Leptons (HNLs).

In this context, muon-specific interactions offer a particularly interesting opportunity since the parameter space for neutral current (NC) four-lepton effective operators involving muons remains largely untested. While previously unconstrained NC operators involving quarks and tau leptons are started to be probed by neutrino oscillation and CE$\nu$NS experiments~\cite{Coloma:2024ict}, some four-lepton operators involving muon and tau flavors remain unbounded by laboratory experiments~\cite{Falkowski:2017pss,Breso-Pla:2023tnz}. If extra beyond SM symmetries inducing strong correlations among different flavors are invoked, those operators can be indirectly constrained (see for instance~\cite{Allwicher:2023shc}). Limits from SN $1987$A cooling on generic effective four-fermion operators that couple muons to long-lived dark fermions, with masses up to $\sim100$ MeV, under reasonable assumptions, have been extracted in~\cite{Manzari:2023gkt}. However, direct laboratory probes have not been considered yet in the literature. The NA64 experiment at CERN has recently published the first results of its muon program, NA64$\mu$, which offers a unique way to probe such operators \cite{NA64:2024klw}. NA64$\mu$ employs a high-energy muon beam combined with a missing energy-momentum technique to search for new light states coupled to muons. In this work, we reinterpret NA64$\mu$ data in the framework of SMEFT and $\nu$SMEFT to derive new bounds on dimension-six four-fermion effective operators involving muons. Our results represent the first direct experimental constraints on a broad class of operators that can be probed via elastic muon-nucleus scattering with missing energy signatures, demonstrating the potential of NA64$\mu$ to probe these interactions. With future upgrades, the total integrated luminosity of the experiment is expected to increase by up to 3 orders of magnitude \cite{NA64:2024nwj}. 

\section{SMEFT and $\nu$SMEFT}
\label{sec:EFT}

New Physics may manifest indirectly in processes occurring at energies below the characteristic scale of the underlying theory. A model-independent way to analyze the indirect effects from New Physics is making use of the EFTs. The power and versatility of EFTs come from the fact that they rely on an operator expansion of dimension higher than 4, that can contemplate all the possibilities regarding the nature of the New Physics interaction. 
If there is no additional low energy New Physics, the EFT should be built upon the SM field content and respect the SM gauge symmetry group. At the same time, the theory should be renormalizable order by order. The resulting EFT is the so-called SMEFT. 
The SMEFT Lagrangian can be written in terms of the dimension of the operators as,
\begin{eqnarray}
    \mathcal{L}= \mathcal{L}_{\rm{SM}} + \mathcal{L}_{d=5} + \mathcal{L}_{d=6}+ ... \,,
\end{eqnarray}
where $\mathcal{L}_{\rm{SM}}$ is the SM Lagrangian 
, and the $d=N$ operators are suppressed by powers $1/\Lambda^{N-4}$ of the new physics scale $\Lambda$. Therefore, beyond the Weinberg $d=5$ operator that leads to neutrino mass generation, the least suppressed New Physics effects are expected to be driven by the $d=6$ contributions, which for convenience we normalize as
\begin{equation}
\label{eq:d6smeft}
    \mathcal{L}_{d=6} = \sum_i \dfrac{c_{i}}{v^2}\mathcal{O}_i\,,
\end{equation}
with $v^2=1/\sqrt{2}G_F$ and $c_{i}$ the Wilson coefficient (WC) that controls the signal strength of each operator $\mathcal{O}_i$. The sum is over all the independent $d=6$ operators that can be built upon the SM particle content and respect its symmetries. In the remainder of this work we will focus our analysis on the impact of the $d=6$ effective operators. 

If the new degrees of freedom have masses that are equal or lower than the characteristic energy transfer of the processes under study, the SMEFT description is no longer valid. In such a case, they should be included in the field content of the low energy effective theory and thus an extension of the SMEFT with these extra building blocks should be considered. In particular, the existence of HNLs below the electroweak scale is well motivated by low scale symmetry protected scenarios~\cite{Branco:1988ex,Kersten:2007vk,Abada:2007ux,Moffat:2017feq}, as the inverse~\cite{Mohapatra:1986aw,Mohapatra:1986bd} or linear~\cite{Akhmedov:1995ip,Malinsky:2005bi} seesaws, able to generate the observed neutrino mass and mixing pattern, the baryon asymmetry of the universe~\cite{Fukugita:1986hr,Akhmedov:1998qx,Asaka:2005pn,Asaka:2005an,Drewes:2017zyw} and Dark Matter abundance~\cite{Asaka:2005pn,Asaka:2005an}. If this is the case, new interactions between the HNL and SM sectors associated to additional heavy physics are efficiently parameterized using the usually referred as $\nu$SMEFT. The new operators thus involve light fermionic degrees of freedom, $N$, with mass $m_N$ that behave as singlets under the SM gauge group. This framework can be extended to encompass effective EFTs relevant to dark matter (DM). By incorporating HNLs or other dark sector particles into the SMEFT, one can systematically construct higher-dimensional operators that describe interactions between DM and SM particles \cite{Aebischer:2022wnl,Baumgart:2022vwr, Liang:2023yta,Song:2023jqm,Song:2023lxf}. In this work, we will focus on $d=6$ operators since higher dimensional operators will be more suppressed and the three $d=5$ operators that can give sizable effects~\cite{Graesser:2007pc,Aparici:2009fh,Caputo:2017pit,Barducci:2020icf,Barducci:2020ncz,Barducci:2022hll,Fernandez-Martinez:2023phj} are not relevant to our study in NA64. 

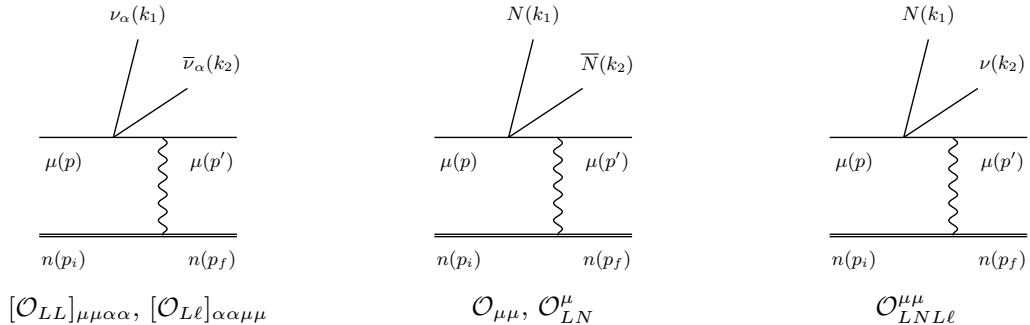
\begin{figure}[t!]
  \begin{center}
    \scalebox{0.65}{
      \begin{tikzpicture}
	\begin{scope}[thick] 
    \draw[thick, -] (0, 4)--(4, 4);
    \draw[thick, double] (0, 2)--(4, 2);
    \draw[thick, photon] (2.5,4)--(2.5,2);
    \draw[thick, -] (1.5,4)--(2,6);
    \draw[thick, -] (1.5,4)--(3,5);
    \node[black,scale=1.25] at (2,6.5) {{$\nu_\alpha (k_1)$}}; 
    \node[black,scale=1.25] at (3.5,5.5) {{$\overline{\nu}_\alpha (k_2)$ }}; 
    \node[black,scale=1.25] at (0.5,3.5) {{$\mu (p)$}};
    \node[black,scale=1.25] at (3.5,3.5) {{$\mu (p')$}};
    \node[black,scale=1.25] at (0.5,1.5) {{$n(p_i)$}};
    \node[black,scale=1.25] at (3.5,1.5) {{$n(p_f)$}};
    \node[black,scale=1.7] at (2.00,0.5) {{$[\mathcal{O}_{LL}]_{\mu\mu\alpha\alpha}$, $[\mathcal{O}_{L\ell}]_{\alpha\alpha \mu\mu}$}};

    \draw[thick, -] (8, 4)--(12, 4);
    \draw[thick, double] (8, 2)--(12, 2);
    \draw[thick, photon] (10.5,4)--(10.5,2);
    \draw[thick, -] (9.5,4)--(10,6);
    \draw[thick, -] (9.5,4)--(11,5);
    \node[black,scale=1.25] at (10,6.5) {{$N (k_1)$}};
    \node[black,scale=1.25] at (11.5,5.5) {{$\overline{N} (k_2)$}};
    \node[black,scale=1.25] at (8.5,3.5) {{$\mu (p)$}};
    \node[black,scale=1.25] at (11.5,3.5) {{$\mu (p')$}};
    \node[black,scale=1.25] at (8.5,1.5) {{$n(p_i)$}};
    \node[black,scale=1.25] at (11.5,1.5) {{$n(p_f)$}};
    \node[black,scale=1.7] at (10.00,0.5) {{$\mathcal{O}_{\mu\mu}$, $\mathcal{O}_{LN}^{\mu}$}};

    \draw[thick, -] (16, 4)--(20, 4);
    \draw[thick, double] (16, 2)--(20, 2);
    \draw[thick, photon] (18.5,4)--(18.5,2);
    \draw[thick, -] (17.5,4)--(18,6);
    \draw[thick, -] (17.5,4)--(19,5);
    \node[black,scale=1.25] at (18,6.5) {{$N (k_1)$}};
    \node[black,scale=1.25] at (19.5,5.5) {{$\nu (k_2)$}};
    \node[black,scale=1.25] at (16.5,3.5) {{$\mu (p)$}};
    \node[black,scale=1.25] at (19.5,3.5) {{$\mu (p')$}};
    \node[black,scale=1.25] at (16.5,1.5) {{$n(p_i)$}};
    \node[black,scale=1.25] at (19.5,1.5) {{$n(p_f)$}};
    \node[black,scale=1.7] at (17.75,0.5) {{$\mathcal{O}_{LNL\ell}^{\mu\mu}$}};
	\end{scope}
      \end{tikzpicture}
    }
  \end{center}
  \caption{Feynman diagrams for the two to four neutrino and HNL production processes at NA64$\mu$ mediated by the indicated effective operators of the SMEFT and $\nu$SMEFT (see Table~\ref{tab:operators}). For simplicity we omit the analogous diagrams in which the photon is interchanged with the nucleus ``before'' the HNL and/or neutrino emission.}
    \label{fig:diag}
\end{figure}

An important part of the SMEFT and $\nu$SMEFT parameter space is already constrained thanks to a plethora of accelerator-based experiments which can probe different combinations of parameters depending on the details of the experimental set up. In particular, LEP provides the leading constraints on four-lepton effective operators involving electrons~\cite{DELPHI:2003dlq,Fox:2011fx,Fernandez-Martinez:2023phj}. However, the four-fermion operators listed in Table~\ref{tab:operators} remain unbounded. NA64$\mu$ offers the unique opportunity to probe indirect New Physics effects encoded in these effective operators via the missing energy signals described in Fig.~\ref{fig:diag}. NA64 running with electrons \cite{NA64:2023wbi} can also probe operators involving electrons but the constraints are weaker than those extracted from LEP or Belle II~\cite{Liang:2021kgw}. 

\begin{table}
\centering
\begin{tabular}{|c|ll|c|c|}
\hline
\multirow{2}{*}{\textbf{Type}} & \multicolumn{2}{c|}{\multirow{2}{*}{\textbf{Operator}}} & \textbf{Current NA64$\mu$} & \textbf{Future NA64$\mu$} \\
 & \multicolumn{2}{c|}{} & \textbf{sensitivity} & \textbf{sensitivity} \\ \hline
\multicolumn{1}{|c|}{\multirow{2}{*}{NC-SMEFT}} & $[\mathcal{O}_{LL}]_{\mu\mu\mu\mu}$   &     $(\overline{L}_\mu\gamma^\mu L_\mu)(\overline{L}_\mu\gamma_\mu L_\mu)$ &        $1.0\cdot 10^{-2}$        &  $1.7\cdot 10^{-4}$ \\ 
\multicolumn{1}{|c|}{} & $[\mathcal{O}_{LL}]_{\mu\mu\tau\tau}$   &     $(\overline{L}_\mu\gamma^\mu L_\mu)(\overline{L}_\tau\gamma_\mu L_\tau)$ &        $1.0\cdot 10^{-2}$        &  $1.3\cdot 10^{-4}$ \\ 
\cline{2-5} 
\multicolumn{1}{|c|}{}                          & $[\mathcal{O}_{L\ell}]_{\mu\mu\mu\mu}$      &       $(\overline{L}_\mu\gamma^\mu L_\mu)(\overline{\ell}_\mu\gamma_\mu \ell_\mu)$ &       $1.0\cdot 10^{-2}$        &      $1.5\cdot 10^{-4}$      \\ 
\multicolumn{1}{|c|}{}                          & $[\mathcal{O}_{L\ell}]_{\tau\tau\mu\mu}$      &       $(\overline{L}_\tau\gamma^\mu L_\tau)(\overline{\ell}_\mu\gamma_\mu \ell_\mu)$ &       $1.0\cdot 10^{-2}$        &      $1.5\cdot 10^{-4}$ \\
\hline
\multirow{2}{*}{NC-$\nu$SMEFT}                  & $\mathcal{O}_{\mu\mu}$ &  $(\overline{\ell}_{\mu}\gamma^\mu \ell_\mu)(\overline{N}\gamma_\mu N)$  &        $4.8\cdot 10^{-3}$                     &         $1.0\cdot 10^{-4}$        \\ \cline{2-5} 
& $\mathcal{O}^\mu_{LN}$ &  $(\overline{L}_\mu \gamma^\mu L_\mu)(\overline{N}\gamma_\mu N)$         &      $4.8\cdot 10^{-3}$          &  $1.0\cdot 10^{-4}$            \\ \hline
CC-$\nu$SMEFT                                   & $\mathcal{O}^{\mu\mu}_{LNL\ell}$ & $(\overline{L}_\mu N)\epsilon(\overline{L}_\mu \ell_\mu)$                    &    $1.5\cdot 10^{-2}$       &       $2.0\cdot 10^{-4}$      \\ \hline
\end{tabular}
\caption{Current and future sensitivity of NA64$\mu$ at 90\% C.L to previously unconstrained four-fermion effective operators in the SMEFT and $\nu$SMEFT frameworks. In the $\nu$SMEFT case the numbers quoted are valid for $m_N\lesssim 2$ GeV.  All bounds are in units of $[{\rm GeV}^{-2}]$.}
\label{tab:operators}
\end{table}

\section{The NA64$\mu$ Experiment}
\label{sec:na64}

\begin{figure}[!h]
\centering
\includegraphics[width=1.\textwidth]{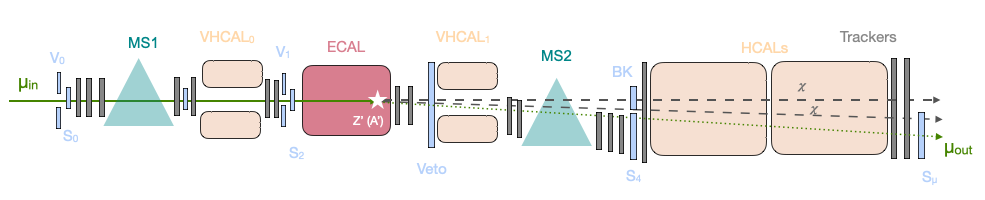}
\caption{Schematic illustration of the NA64$\mu$ set-up in pre-LS3 runs. The spectrometer in the upstream region (MS1) is used for identifying incoming muons with momentum $p_\text{in}\simeq160$ GeV. The downstream part composed of calorimeters and a second spectrometer (MS2) measures the momentum of the scattered muons in the ECAL serving as an active target to search for the production of new light feebly interacting particles.}
\label{fig:NA64_setup}
\end{figure}

The NA64$\mu$ experiment utilizes the M2 beamline at the CERN Super Proton Synchrotron (SPS) accelerator which delivers a 160~GeV muon beam to an active target, an electromagnetic calorimeter (ECAL), followed by a hermetic detector system. A sketch of the setup including an example of a signal-like event is illustrated in Fig. \ref{fig:NA64_setup}. The signal signature is a single scattered muon with momentum below 80~GeV and an energy deposition compatible with a minimum ionizing particle in ECAL and in the hadronic calorimeters (HCAL0-1) located downstream. In addition, it is required no additional energy associated to electromagnetic or hadronic activity in the VETO and the Veto hadronic calorimeters, (VHCAL0-1) downstream the ECAL. If such an event would be detected it would indicate the possible production of invisible states via processes like $\mu \mathcal{N} \rightarrow \mu \mathcal{N} X$, where $\mathcal{N}$ is the target nucleus, and $X$ is a light feebly interacting particle such as a dark photon or a dark Z. In the NA64 kinematic regime, the momentum transfer to the nucleus, and hence the recoil, is negligible. Consequently, the nucleus can be treated as a coherent object, leading to an enhancement of elastic scattering, while inelastic processes are suppressed~\cite{Kirpichnikov:2021jev}.

The incoming muons are tagged by a set of scintillator counters ($S_{0-3}$ and $\overline{V}_{0-1}$) and their momentum is reconstructed by the first magnetic spectrometer (MS1), with Micromegas trackers. The target is a high-granularity Lead-Scintillator ECAL with 40 radiation length, capable of measuring energy deposits and identifying minimum-ionizing particles. The scattering of the final state muon is tagged by two counters placed along the beam-deflected axis, $S_4$ and S$_{\mu}$ and its outgoing momentum is reconstructed in the downstream magnetic spectrometer (MS2) with a set of GEMs, Micromegas and Straw Chamber detectors.
The dedicated set of veto and trigger scintillators ensures rejection of background processes such as kaon decays, beam halo interactions, and mis-reconstructed tracks. The HCAL modules close the setup hermeticity together with the veto detectors. 

For the 2022 dataset of $1.98 \cdot 10^{10}$ muons on target (MOT), no candidate events were observed, allowing NA64 to place the most stringent constraints to date on light vector mediators coupling to muons with the $L_\mu - L_\tau$ gauge extension of the SM as a benchmark model \cite{NA64:2024klw}. In 2023/2024 runs $3.5\cdot 10^{11}$ MOT were collected and the analysis is ongoing. The experiment could potentially reach $1.0\cdot10^{14}$ MOT during LHC Run 5. An upgrade of the experimental setup is currently being studied by the collaboration to reach such milestone. 

\section{Muon four-fermion effective operators at NA64}
\label{sec:4fermion}

In this section we will introduce the effective operators from the SMEFT and $\nu$SMEFT that can be probed by NA64$\mu$. We will also briefly clarify which are the operators that currently can only be constrained by NA64$\mu$. The list with the most relevant operators is given in Table~\ref{tab:operators}. 

\subsection{SMEFT}
\label{sec:SMEFT}
Given that the characteristic momentum transfer at NA64$\mu$ is much smaller than the electroweak scale, the New Physics effects associated to the existence of new and heavier BSM physics can be conveniently parameterized by the so called Weak Effective Field Theory (WEFT)~\cite{Jenkins:2017jig}. In this EFT not only the BSM degrees of freedom but also the SM $W$, $Z$ and Higgs bosons, are integrated out. The new interactions are thus encoded in four-fermion effective operators. In this framework, the relevant operators that can be probed by NA64$\mu$ are the following 
\begin{equation}
    \mathcal{L}_{\text{WEFT}}\supset 
    -\sqrt{2}G_F \varepsilon^{\mu,V}_{\alpha\beta}(\overline{\nu}_\alpha \gamma_\mu P_L\nu_\beta)(\overline{\mu}\gamma^{\mu} \mu)+\sqrt{2}G_F \varepsilon^{\mu,A}_{\alpha\beta}(\overline{\nu}_\alpha \gamma_\mu P_L\nu_\beta)(\overline{\mu}\gamma^\mu\gamma_5\mu)\,.
    \label{eq:WEFT}
\end{equation}
Considering the flavor conserving case, usually assumed in global SMEFT analysis, the Wilson coefficients $\varepsilon^{\mu,V}_{\alpha\alpha}$ and $\varepsilon^{\mu,A}_{\alpha\alpha}$ can be mapped to those of the $d=6$ SMEFT operators~\cite{Falkowski:2017pss}
\begin{eqnarray}
\varepsilon^{\mu,V}_{\alpha\alpha}&=&  \delta_{\mu\alpha}\left(\delta g_L^{W\mu}- \delta g_L^{W e} + \frac{1}{2}[c_{LL}]_{e\mu\mu e}\right)
    -(1-4s_{\rm w}^2)\delta g_L^{Z\nu_\alpha}+\delta g_L^{Z\mu}+\delta g_R^{Z\mu} - \nonumber\\ 
  &-& \frac{1}{2}\Big(x_{\mu\alpha}+[c_{L\ell}]_{\alpha\alpha \mu\mu}\Big)\,,
    \\
    \varepsilon^{\mu,A}_{\alpha\alpha}&=& \delta_{\mu\alpha}\left(\delta g_L^{W\mu}- \delta g_L^{W e} + \frac{1}{2}[c_{LL}]_{e\mu\mu e}\right)
    -\delta g_L^{Z\nu_\alpha}+\delta g_L^{Z\mu}-\delta g_R^{Z\mu}-\frac{1}{2}\Big(x_{\mu\alpha}-[c_{L\ell}]_{\alpha\alpha \mu\mu}\Big)\,,  \nonumber
    \label{eq:weft}
\end{eqnarray}
where $\delta g^W$ and $\delta g^Z$ are the SMEFT corrections to the vertex between the fermions and the corresponding gauge bosons, 
$x_{\mu\alpha}=[c_{LL}]_{\mu\mu\alpha\alpha}$ for $\alpha=\mu,\tau$ and $x_{\mu e}=[c_{LL}]_{e e \mu\mu}$. 
The coefficients $[c_{LL}]_{\mu\mu \alpha\alpha}$ and $[c_{L\ell}]_{\alpha\alpha \mu\mu}$ are the WC of the operators
\begin{eqnarray}
\left[\mathcal{O}_{LL}\right]_{\mu\mu \alpha\alpha}&=&\dfrac{[c_{LL}]_{{\mu\mu \alpha\alpha}}}{v^2}(\overline{L}_\mu\gamma^\nu L_\mu)(\overline{L}_\alpha\gamma_\nu L_\alpha)\,,
\nonumber \\
 \left[\mathcal{O}_{L\ell}\right]_{{\alpha\alpha \mu\mu}} &=&\dfrac{[c_{L\ell}]_{{\alpha\alpha \mu\mu}}}{v^2}(\overline{L}_\alpha\gamma^\nu L_\alpha)(\overline{\ell}_\mu\gamma_\nu \ell_\mu)\,,
 \label{eq:Osmeft}
\end{eqnarray}
%
respectively, normalized as in Eq.~(\ref{eq:d6smeft}) and with $L$ ($\ell$) denoting $SU(2)$ lepton doublet (singlet).\footnote{Notice that in some SMEFT literature the commonly used notation for $SU(2)$ lepton doublet (singlet) is $\ell$ ($e$), see for instance~\cite{Falkowski:2017pss,Breso-Pla:2023tnz}.} 

NA64$\mu$ is sensitive to a linear combination of the vectorial and axial effective couplings
\begin{eqnarray}
\sum_\alpha \left(a\,\left| \delta_{\alpha\mu}-\frac{1}{2}+2s_{\rm w}^2+\varepsilon^{\mu,V}_{\alpha\alpha}\right|^2+b\,\left| \delta_{\alpha\mu}-\frac{1}{2}+\varepsilon^{\mu,A}_{\alpha\alpha}\right|^2\right),
\label{eq:sumepsilon}
\end{eqnarray}
which ultimately depends on the SMEFT parameters shown in Eq.~\eqref{eq:weft}. However, most of the SMEFT parameters involved hold current bounds~\cite{Falkowski:2017pss,Breso-Pla:2023tnz} that are at least two orders of magnitude stronger than the NA64$\mu$ future sensitivity with the notable exception of $[c_{LL}]_{\mu\mu\tau\tau}$ and $[c_{L\ell}]_{\tau\tau \mu\mu}$ which, to our knowledge, are currently unbounded. Additionally, only the following combination of the parameters $[c_{LL}]_{\mu\mu\mu\mu}$ and $[c_{L\ell}]_{\mu\mu\mu\mu}$
\begin{eqnarray}
[\hat{c}_{LL}]_{\mu\mu\mu\mu}=[c_{LL}]_{\mu\mu\mu\mu}+\dfrac{2 g_Y^2}{g_L^2+3g_Y^2}[c_{L\ell}]_{\mu\mu\mu\mu}\,,
\label{eq:flatdirection}
\end{eqnarray}
is currently constrained~\cite{Falkowski:2017pss}. Therefore, we can perform our analysis in terms of $[c_{LL}]_{\mu\mu\tau\tau}$, $[c_{L\ell}]_{\tau\tau \mu\mu}$, $[c_{LL}]_{\mu\mu\mu\mu}$ and $[c_{L\ell}]_{\mu\mu\mu\mu}$, and safely neglect the rest SMEFT parameters considering
\begin{eqnarray}
\varepsilon^{\mu,V}_{\alpha\alpha}&=&  - \frac{1}{2}\Big([c_{LL}]_{\mu\mu\alpha\alpha}+[c_{L\ell}]_{\alpha\alpha\mu\mu}\Big)\,,
    \nonumber\\
 \varepsilon^{\mu,A}_{\alpha\alpha}&=& -\frac{1}{2}\Big([c_{LL}]_{\mu\mu\alpha\alpha}-[c_{L\ell}]_{\alpha\alpha\mu\mu}\Big)\,,
    \\
\varepsilon^{\mu,V}_{ee}&=&\varepsilon^{\mu,A}_{ee}=0\,, \nonumber
    \label{eq:weftNA64}
\end{eqnarray}
with $\alpha=\mu,\tau$. In other words, NA64$\mu$ can probe the unconstrained effective operators $\left[\mathcal{O}_{LL}\right]_{\mu\mu\tau\tau}$ and $\left[\mathcal{O}_{L\ell}\right]_{\tau\tau\mu\mu}$, and potentially break the flat direction defined by  Eq.~(\ref{eq:flatdirection}).

%



\subsection{$\nu$SMEFT}

NA64$\mu$ can probe two NC-like $\nu$SMEFT operators involving leptons:
\begin{eqnarray}
 \mathcal{O}_{\mu\mu}&=&\dfrac{C_{\mu\mu}}{\Lambda^2}(\overline{\ell}_\mu\gamma^\mu \ell_\mu)(\overline{N}\gamma_\mu N)\,,
 \label{eq:Omumu}
 \\
 \mathcal{O}_{LN}^\mu &=&\dfrac{C_{LN}^\mu}{\Lambda^2}(\overline{L}_\mu\gamma^\mu L_\mu)(\overline{N}\gamma_\mu N)\,,
 \label{eq:OllOle} 
\end{eqnarray}
where the first operator involves only singlets of $SU(2)$ (two right-handed leptons, $\ell_\mu$, and two singlet fermions, $N$), while the second involves two lepton doublets $L_\mu$ and two fermion singlets. And the following CC-like operator
\begin{eqnarray}
    \mathcal{O}^{\mu\mu}_{LNL\ell}=\dfrac{C_{LNL\mu}^{\mu\mu}}{\Lambda^2}(\overline{L}_\mu N)\epsilon(\overline{L}_\mu \ell_\mu)\,.
       \label{eq:OLN}
\end{eqnarray}

To our knowledge, so far the missing energy/momentum signal of NA64$\mu$ is the only observable sensitive to these operators. Conversely, the equivalent operators involving electrons are strongly constrained by LEP monophoton searches~\cite{DELPHI:2003dlq,Fox:2011fx,Fernandez-Martinez:2023phj} and supernova cooling~\cite{DeRocco:2019jti,Fernandez-Martinez:2023phj}. 





\section{Analysis and results}
\label{sec:results}

The relevant process at NA64$\mu$ is the following
\begin{eqnarray}
    \mu(p) + \mathcal{N}(p_i) \to \mu (p^\prime) + \mathcal{N}(p_f) + \chi_1(k_1) + \chi_2 (k_2)\,,
\end{eqnarray}
where $\mathcal{N}$  here represents the nucleus of lead, we specify the momentum carried out by each particle and denote $\chi_1$ and $\chi_2$ as generic neutral fermions which in our case will be the HNLs or SM neutrinos depending on the effective operator mediating the process (see Fig.~\ref{fig:diag} and Table~\ref{tab:operators}). The signal is the significant amount of missing energy carried out by the neutral particles.

The key quantity to extract the current NA64$\mu$ bound and future sensitivity to the SMEFT and $\nu$SMEFT is the production cross section of this process, which can be conveniently separated into the convolution of a $1\rightarrow2$ and a $2\rightarrow 3$ process~\cite{Gninenko:2018ter,Liang:2021kgw}
\begin{eqnarray}
\frac{d\sigma (\mu \mathcal{N}\to\mu \mathcal{N} \chi_1\chi_2)}{dk^2}&=&\sum_{i,j}\Bigl\{\frac{1}{2\pi} \int d\Phi_2(k_1,k_2) \sum_{s_1,s_2} \mathcal{J}_i^\mu(\mathcal{J}_j^\nu)^\dagger\Bigr\}\times\\
&\times&\Bigl\{\int d\Phi_3(p_f,p^\prime,k) \frac{\overline{\mathcal{M}_{i\mu} \mathcal{M}_{j\nu}^\dagger}}{4|\vec{p}|M} \Bigr\}\,. \nonumber
\end{eqnarray}
where $\mathcal{J}_i^{\mu}(k_1,k_2)=\overline{u}(k_1)\Gamma_i^\mu v(k_2)$ is the fermion current of the outgoing $\chi_1$ and $\chi_2$, with $\Gamma^\mu_i$ the Lorentz structure of the current, $\mathcal{M}_{j\nu}(p,p^\prime,p_i,p_f,k)$ is the amplitude of the remaining $2\rightarrow 3$ process with $k=k_1+k_2$. The $i,j$ indices indicate the sum over the operators mediating the process. The factorization of the $2\to4$ phase space is obtained by introducing an intermediate momentum of a virtual particle ($k=k_1+k_2$) and inserting a representation of unity that enforces, through Dirac delta distributions, that $k^\mu=k_1^\mu+k_2^\mu$ and $k^2=(k_1+k_2)^2$. This allows the $2\to 4$ phase space to be written as a convolution of a $2\to3$ phase space, describing the production process $p+p_i \to p' + p_f +k$, and a $1\to2$ phase space, corresponding to the decay of the off-shell virtual particle $k\to k_1 +k_2$. In this way, the four-body kinematics are organized into the convolution of a sequential production–decay process. The integration over the invariant mass $k^2$ of the intermediate virtual state links the two contributions. 
For the SMEFT case, following the spirit of the global analyses, we will simultaneously consider the contribution of the four relevant operators as discussed in Sec.~\ref{sec:SMEFT}. In the $\nu$SMEFT case we will introduce the operators one at a time and therefore we will consider $i=j$ and no sum over operators. The SM contribution should always be considered in both cases; however, for notational convenience, we will omit it, as well as the operator indices $i,j$, in the remainder of this section (see Eqs.~(\ref{eq:sigmaSMEFT}), (\ref{eq:sigmanuSMEFTnc}) and (\ref{eq:sigmanuSMEFTcc}) for the complete final expressions).

We can rearrange the differential cross section and write~\cite{Gninenko:2018ter,Liang:2021kgw},
\begin{eqnarray}
    d\sigma (\mu \mathcal{N}\to \mu \mathcal{N}\chi_1 \chi_2)=d\sigma^{2\to 3}_{\mu\nu}\left|\frac{c}{\Lambda^2}\right|^2 \frac{dk^2}{(2\pi)}\xi^{\mu\nu}\,,
\end{eqnarray}
where  
\begin{eqnarray}
\xi^{\mu\nu}=\int d\Phi_2(k_1,k_2)\sum_{\rm spins}\mathcal{J}^\mu(\mathcal{J}^\nu)^\dagger\,,
\end{eqnarray}
is the $1\to2$ process integrated over the 2-body phase space and $d\sigma^{2\to 3}_{\mu\nu}=d{\rm Lips}_{2\to 3}|\overline{\mathcal{M}_{2\to 3}}|^2_{\mu\nu}$, where we define $d{\rm Lips}_{2\to 3}$ as the $2\to 3$ Lorentz invariant phase space and the squared matrix element averaged over initial lepton spin as 
\begin{eqnarray}
    |\overline{\mathcal{M}_{2\to 3}}|^2_{\mu\nu}\equiv\sum_{\rm spins} \mathcal{M}_\mu\mathcal{M}^\dagger_\nu \,.
\end{eqnarray}
\begin{figure}[t!]
  \begin{center}
    \scalebox{0.8}{
      \begin{tikzpicture}
	\begin{scope}[thick] 
    \draw[thick, -] (0, 4)--(4, 4);
    \draw[thick, photon] (0, 2)--(2.5,4);
    \draw[thick, -] (1.5,4)--(2,6);
    \draw[thick, -] (1.5,4)--(3,5);
    \node[black,scale=1.25] at (2,6.5) {{$\chi_1 (k_1)$}}; 
    \node[black,scale=1.25] at (3.5,5.5) {{$\chi_2 (k_2)$ }}; 
    \node[black,scale=1.25] at (-0.5,4) {{$\mu (p)$}};
    \node[black,scale=1.25] at (4.5,4) {{$\mu (p')$}};
    \node[black,scale=1.25] at (2,2.5) {{$\gamma (q)$}};


    \draw[thick, -] (8, 4)--(12, 4);
    \draw[thick, photon] (8,2)--(9.5,4);
    \draw[thick, -] (10.5,4)--(11,6);
    \draw[thick, -] (10.5,4)--(12,5);
    \node[black,scale=1.25] at (11,6.5) {{$\chi_1 (k_1)$}};
    \node[black,scale=1.25] at (12.5,5.5) {{$\chi_2 (k_2)$}};
    \node[black,scale=1.25] at (7.5,4) {{$\mu (p)$}};
    \node[black,scale=1.25] at (12.5,4) {{$\mu (p')$}};
    \node[black,scale=1.25] at (9.5,2.5) {{$\gamma (q)$}};


	\end{scope}
      \end{tikzpicture}
    }
  \end{center}
  \caption{Feynman diagrams contributing to neutral fermion production in the EFT at NA64$\mu$ using the WW approximation. We denote $\chi_1$ and $\chi_2$ as generic neutral fermions (in our case SM neutrinos or HNLs depending on the effective operator mediating the process).}
    \label{fig:WWA}
\end{figure}
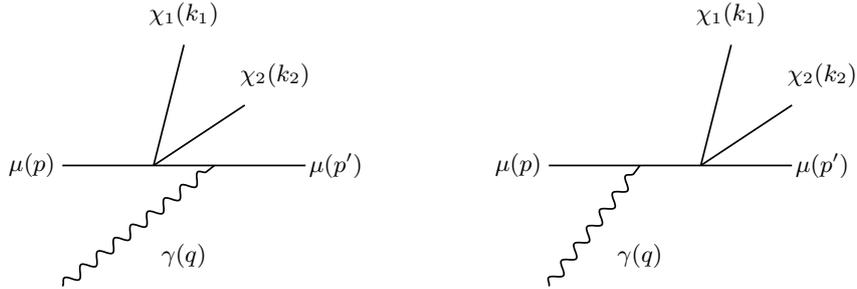
The expressions for $\xi^{\mu\nu}$ can be computed by direct integration. In order to compute the $2\to 3$ process, $d\sigma^{2\to 3}_{\mu\nu}$, we make use of the Weisz\"acker-William (WW) approximation \cite{Kim:1973he, Gninenko:2017yus}. Under this approximation the interaction of the muon with the nucleus can be 
treated as the interaction with an effective photon flux as we show in Fig.~\ref{fig:WWA}. The differential cross section of the $2\to3$ process reads~\cite{Gninenko:2018ter, Kirpichnikov:2021jev, Liang:2021kgw,Sieber:2024vhx,NA64:2024nwj},
\begin{eqnarray}
    \left.\frac{d\sigma_{\mu\nu}^{2\to 3}}{dx}\right|_{\rm WW} = \frac{\alpha}{16\pi^2}\frac{1-x}{x}\sqrt{x^2-\frac{k^2}{E_\mu^2}}\int_{\tilde{u}_{\rm min}}^{\tilde{u}_{\rm max}}\frac{d\tilde{u}}{\tilde{u}^2}|\overline{\mathcal{M}_\mu\mathcal{M}^\dagger_\nu }|_{2\to 3}\chi^{\rm WW},
\end{eqnarray}
where $x=k^0/E_\mu$ is the fraction of the muon energy transferred to the $\chi_1$, $\chi_2$ pair
, $\tilde{u}$ is a Mandelstam variable defined as $\tilde{u}=(p-k)^2-m_\mu^2\simeq-E_\mu^2\theta_k^2x-k^2(1-x)/x-m_\mu^2x$, where $\theta_k$ is the angle of emission of the $\chi_1\chi_2$ pair. The integration limits are given by $\tilde{u}_{\rm min}=\tilde{u}(\theta_k=0.1)$ and $\tilde{u}_{\rm max}=\tilde{u}(\theta_k=0)$~\cite{Kirpichnikov:2021jev, Sieber:2024vhx}. The photon flux, $\chi^{\rm WW}$, is defined as~\cite{Kirpichnikov:2021jev},
\begin{eqnarray}
    \chi^{\rm WW}=\int_{t_{\rm min}}^{t_{\rm max}}dt \frac{t-t_{\rm min}}{t^2} \mathcal{F}^2(t)=Z^2\int_{t_{\rm min}}^{t_{\rm max}} dt \frac{t-t_{\rm min}}{t^2}\left(\frac{t}{t_a+t}\right)^2\left(\frac{t_d}{t_d +t}\right)^2,
\end{eqnarray}
where $\mathcal{F}(t)$ is the nucleus-photon form factor, $Z_{\rm Pb}=82$, $\sqrt{t_a}\simeq2.0\cdot10^{-5}$~GeV, $\sqrt{t_d}\simeq 6.9\cdot10^{-2}$~GeV. Here the integration limits\footnote{Note that we use the WW approximation where the photon flux depends on $\tilde{u}$ and cannot be taken out of the integral. The simplified case in which the integration limit is set to $t_{\rm min}=k^4/4E^2_\mu$ in order to factorize the flux out of the integral is referred as Improved WW (IWW) approximation~\cite{Kim:1973he, Kirpichnikov:2021jev}.  Even though the IWW approximation simplifies calculations and permits to use less computing time, due to the simplification of the photon flux it leads to less accurate results in comparison with the WW approximation that we consider in this work.} are $t_{\rm min}=\tilde{u}^2/(4E_\mu^2(1-x)^2)$ and $t_{\rm max}=m_\mu^2+k^2$.
In the WW regime, the dominant contribution to the cross section arises from the region of small momentum transfer $t$, close to $t_{\rm min}$. In this limit, and within the kinematic range considered in this work, the typical momentum transfer is $\sqrt{|t|}\sim \mathcal{O}(1\text{--}30)\ \text{MeV}$. This scale is comparable to or smaller than the inverse nuclear radius of lead, $1/R_{\rm Pb}\sim 30\ \text{MeV}$. Consequently, the associated wavelength is larger than the nuclear size, and the interaction occurs in the coherent regime, as highlighted in Section~\ref{sec:na64}.

The differential cross section then reads,
\begin{eqnarray}
   \frac{d\sigma(\mu \mathcal{N}\to\mu \mathcal{N} \overline{\chi}\chi)}{dx}= \left|\frac{c}{\Lambda^2}\right|^2 \frac{dk^2}{2\pi}\frac{\alpha^2}{4\pi}\frac{1-x}{x}\sqrt{x^2-\frac{k^2}{E_\mu^2}}\int_{\tilde{u}_{\rm min}}^{\tilde{u}_{\rm max}}\frac{d\tilde{u}}{\tilde{u}^2}\xi^{\mu\nu}_{1\to 2}|\overline{\mathcal{A}^{2\to 2}}|_{\mu\nu}^2\,\,\chi^{\rm WW}\,,
\end{eqnarray}
where we have defined $|\overline{\mathcal{M}_\mu\mathcal{M}^\dagger_\nu }|_{2\to 3}=e^2|\overline{\mathcal{A}^{2\to 2}}|_{\mu\nu}^2$~\cite{Kirpichnikov:2021jev,Sieber:2024vhx}. In order to handle better the integral one can express the result of contracting the amplitudes as 
\begin{eqnarray}
    \frac{d\sigma}{dxdk^2}= \frac{\alpha^2}{48\pi^3}\left|\frac{c}{\Lambda^2}\right|^2
    \frac{1-x}{x}\sqrt{x^2-\frac{k^2}{E_\mu^2}}\beta(k,m_{\chi_1},m_{\chi_2})\int \frac{d\tilde{u}}{\tilde{u}^2}\,\,\chi^{\rm WW}\sum_{n=1}^3\frac{\mathcal{C}_n}{\tilde{u}^{n-1}}\,,
    \label{eq:dsdxdk2}
\end{eqnarray}
where the coefficients $\mathcal{C}_n$ for the different cases under consideration are given in Appendix~\ref{app:XS}, 
and the function $\beta(k,m_1,m_2)$ is given by
\begin{eqnarray}
\beta(k,m_{1},m_{2})=\sqrt{\left(1-\frac{m_1^2+m_2^2}{k^2}\right)^2-\frac{4m_1^2m_2^2}{k^4}} \,.
\label{eq:phase_space}
\end{eqnarray}

\begin{figure}
	\centering
        \includegraphics[width=.49\textwidth]{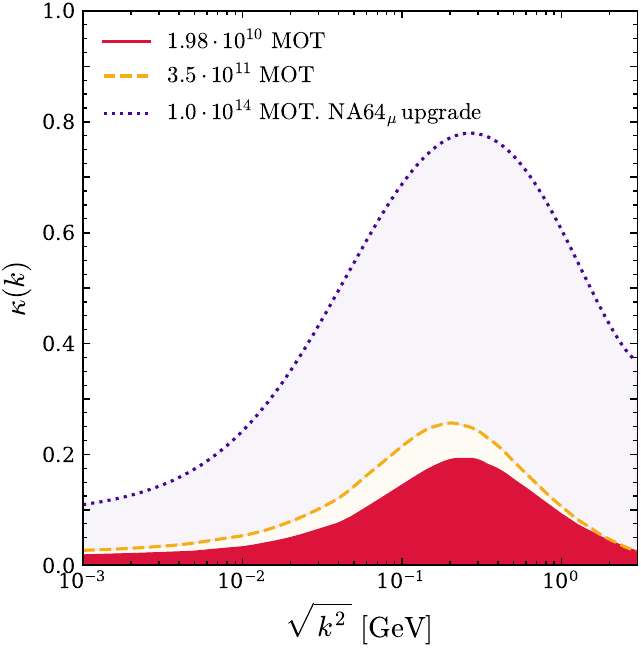}
        \caption{
        Geometrical acceptance on the signal efficiency as a function of $k^2$. The magenta solid line is the acceptance for published data extracted from~\cite{NA64:2024nwj}. The yellow dashed line and the purple dotted line indicate the expected curve for the ongoing analysis, and the future expected acceptance respectively.}
	\label{fig:acc}
\end{figure}

Finally, in order to compute the total cross section and obtain the number of NA64$\mu$ signal events it is necessary to perform the integral over $k^2$ and $x$~\cite{NA64:2024nwj}
\begin{eqnarray}
    N_{S}= N_{\rm MOT}\frac{\rho_{\mathcal{N}}}{m_{\mathcal{N}}}L_{\rm T}^{\rm eff}\int_{(m_{\chi_1}+m_{\chi_2})/E_\mu}^{1-\frac{m_\mu}{E_\mu}}dx\int_{(m_{\chi_1}+m_{\chi_2})^2}^{x^2E_\mu^2}dk^2 \kappa (k) \frac{d\sigma}{dxdk^2}\,,
    \label{eq:nevents}
\end{eqnarray}
where $N_{\rm MOT}$ is the number of muons on target, $\rho_{\mathcal{N}}$ and $m_{\mathcal{N}}$ are the density and the nucleus mass of the target material respectively and $L_{\rm T}^{\rm eff}$ is the effective thickness of the target material. The target material used in NA64 is lead with $\rho_{\rm Pb}=11.35$~g~cm$^{-3}$ and $m_{\rm Pb}\simeq 207$~GeV and the effective thickness is $L_{\rm Pb}^{\rm eff}\simeq~22.5$~cm. After the NA64 upgrade, the effective target length is expected to be three times longer. The effect of the geometrical acceptance and efficiency is parametrized by the function $\kappa(k)$.
In Fig.~\ref{fig:acc}, the acceptance as a function of $k$, computed in Ref.~\cite{NA64:2024nwj}, is shown as a magenta solid line for the published data set corresponding to $1.98\cdot10^{10}$~MOT. During the 2023–2024 runs, $3.5\cdot10^{11}$~MOT were collected with an improved setup, whose acceptance is indicated by the yellow dashed line. Finally, for future data-taking periods after the next CERN shutdown, NA64$\mu$ setup will be upgraded. The purple line shows the expected geometrical acceptance for those runs.
The acceptance is obtained for models in which a dark photon $Z'$ is the mediator of the process and depends on the dark photon mass $m_{Z'}$. Nevertheless, since NA64 is only sensitive to the missing energy/momentum or the invisible $Z'$ decay, in our EFT case the acceptance is a function of $k^2=(k_1+k_2)^2$.

In order to set limits to the different effective operators we make use of Eq.~\eqref{eq:nevents} to compute the expected number of signal events for each operator (or combination of operators in the SMEFT case) and require $N_S\leq 2.44$ that corresponds to the 90\% confidence level (CL) exclusion limit assuming zero background and null event result in the data~\cite{Feldman:1997qc}. 


\begin{figure}
	\centering
        \includegraphics[width=.49\textwidth]{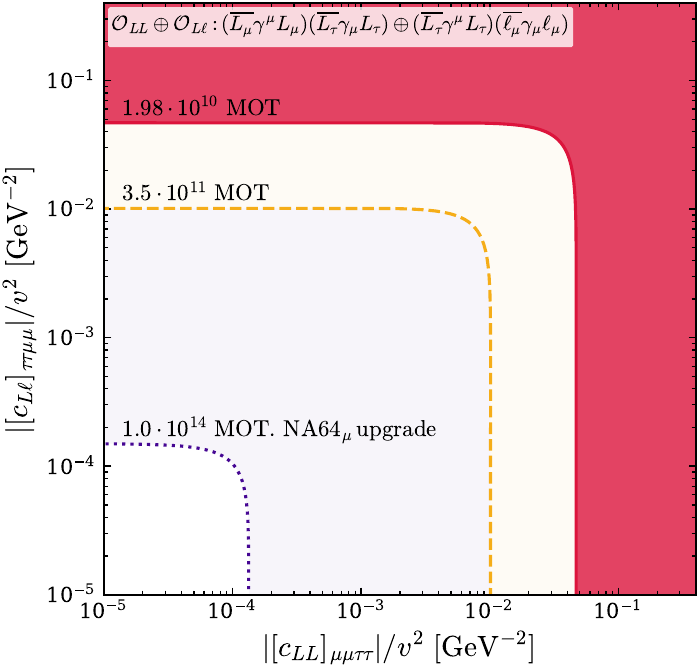}
        \caption{90\% CL NA64$\mu$ sensitivity to the WC of the SMEFT operators given in Eq.~(\ref{eq:Osmeft}) with $\alpha=\tau$ and the three data sets introduced in Sec.~\ref{sec:na64}. We omit here the results for $\alpha=\mu$ since they are extremely similar to the $\alpha=\tau$ case.
     }
	\label{fig:resultssmeft}
\end{figure}

In the SMEFT scenario we introduce simultaneously four effective operators whose effect is encoded in the effective parameters given in Eq.~\eqref{eq:Osmeft}. As we have already mentioned, NA64$\mu$ is sensitive to a combination of these parameters (see Eq.~\eqref{eq:sumepsilon} and the expression for the production cross section in Appendix~\ref{app:XS}). Our results are presented in Fig.~\ref{fig:resultssmeft} for the unbounded operators with $\alpha=\tau$, where we show the region of the SMEFT parameter space already excluded by NA64$\mu$ (magenta shaded area), the sensitivity using the 2023-2024 collected statistics (yellow dashed line), and the expected future sensitivity (purple dotted line) as indicated in the figure. We do not show the results for $\alpha=\mu$ since we obtain essentially the same current bound and future sensitivity (see Table~\ref{tab:operators}). 
The projected sensitivities shown in Fig.~\ref{fig:resultssmeft} corresponds to an upper bound on the WCs of the SMEFT operators discussed in Sec.~\ref{sec:SMEFT}, given by 
\begin{eqnarray}
    \left|\frac{[c_{LL}]_{\mu\mu\alpha\alpha}}{v^2}\right|^2 + \left|\frac{[c_{L\ell}]_{\alpha\alpha\mu\mu}}{v^2}\right|^2 - 0.1 \left|\frac{[c_{LL}]_{\mu\mu\alpha\alpha}[c_{L\ell}]_{\alpha\alpha\mu\mu}}{v^4} \right| - \mathcal{A}\lesssim& 2.0\cdot 10^{-8}~\text{GeV}^{-4}\,,
\label{eq:SMEFT_bound}
\end{eqnarray}
with the interference between SM and SMEFT given by $\mathcal{A} = 1.2\cdot 10^{-5} |[c_{LL}]_{\mu\mu\mu\mu}/v^2| + 4.7\cdot~10^{-5} |[c_{Ll}]_{\mu\mu\mu\mu}/v^2|$ or $\mathcal{A} = -1.9\cdot 10^{-5} |[c_{LL}]_{\mu\mu\tau\tau}/v^2| + 1.6\cdot 10^{-5} |[c_{Ll}]_{\tau\tau\mu\mu}/v^2|$, for $\alpha=\mu$ and $\alpha=\tau$, respectively.
Moreover, in the case of $\alpha=\mu$, the flat direction among four-muon operators in the trident observable, given in Eq.~\eqref{eq:flatdirection}, is constrained to be $[\hat c_{LL}]_{\mu\mu\mu\mu}\leq 0.21$ (68\%~CL) by the global fit in Ref.~\cite{Falkowski:2017pss,Breso-Pla:2023tnz}. The combination of both constraints results in the following future combined sensitivity at 90\%~CL 
\begin{eqnarray}
    \left|\frac{[c_{LL}]_{\mu\mu\mu\mu}}{v^2}\right| \lesssim 4.8\cdot 10^{-5}~\text{GeV}^{-2} \quad \text{and} \quad \left|\frac{[c_{L\ell}]_{\mu\mu\mu\mu}}{v^2}\right| \lesssim 1.4\cdot 10^{-4}~\text{GeV}^{-2}\,.
\label{eq:SMEFT_combined_bound}
\end{eqnarray}

NA64$\mu$ can thus break the flat direction defined by Eq.~\eqref{eq:flatdirection}. Note also that the combined sensitivity slightly improves over the NA64$\mu$ alone sensitivity shown in Fig.~\ref{fig:resultssmeft}.

\begin{figure}
	\centering
        \includegraphics[width=.49\textwidth]{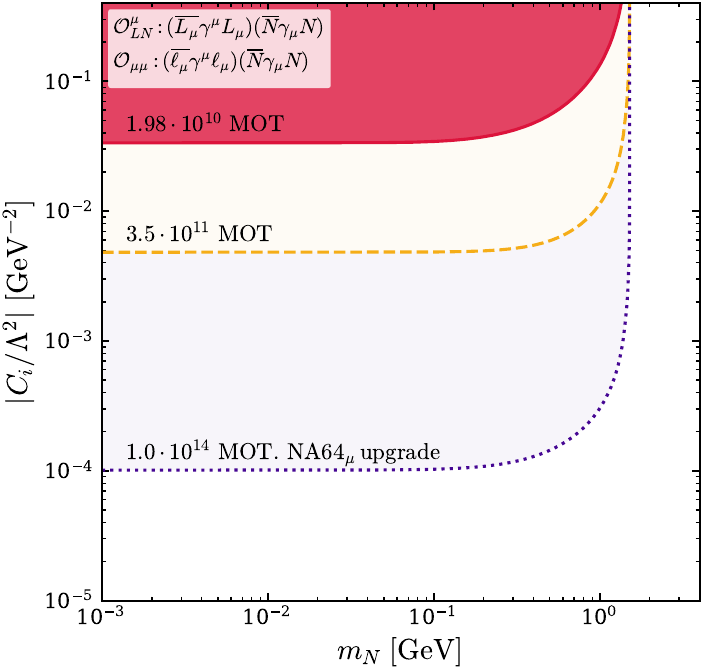}
        \includegraphics[width=.49\textwidth]{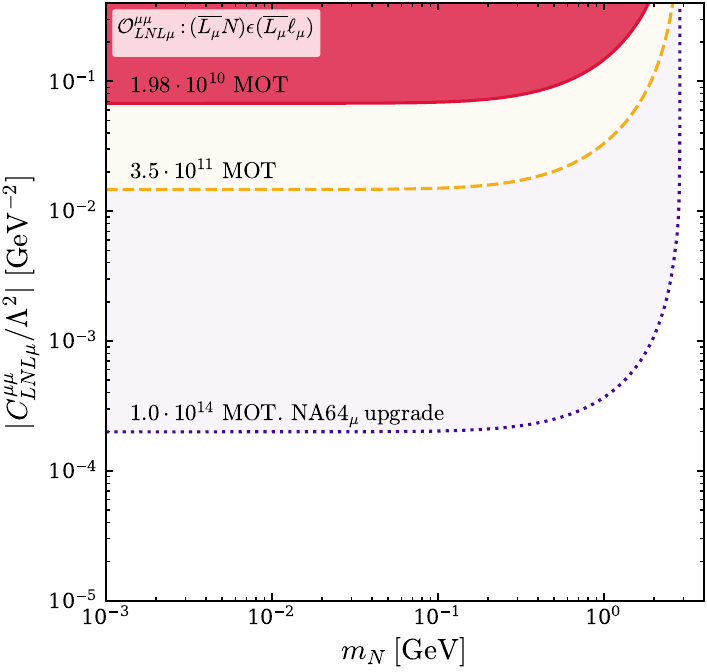}
        \caption{90\% CL NA64$\mu$ current bound and expected sensitivities to the WC of the $\nu$SMEFT operators given in Eqs.~(\ref{eq:Omumu})-(\ref{eq:OLN}).}
	\label{fig:results}
\end{figure}

The NA64$\mu$ sensitivity to the $\nu$SMEFT is shown in Fig.~\ref{fig:results}, using the same color coding as in Fig.~\ref{fig:resultssmeft}. Here, solid, dashed, and dotted lines correspond to the excluded region, current sensitivity, and expected future sensitivity, respectively, as in the previous figure.  
The limit $m_N\lesssim 1.5~(3)$~GeV for the effective operators $\mathcal{O}_{\mu\mu}$ and $\mathcal{O}^\mu_{LN}$ ($\mathcal{O}^{\mu\mu}_{LNL\ell}$) comes from the closing of the phase space in the cross section given by Eq.~(\ref{eq:phase_space}) when the geometric acceptance reaches the highest Monte Carlo simulated values of $\sqrt{k^2}$ in Fig.~\ref{fig:acc}. As expected, qualitatively we obtain similar bounds and future sensitivities for all the operators. The slight differences between the sensitivities to the different SMEFT and $\nu$SMEFT operators arise mainly due to the different chiral structures.



\section{Outlook and conclusions}
\label{sec:conclusions}

The reinterpretation of NA64$\mu$ results in terms of SMEFT and $\nu$SMEFT operators provides the first experimental bounds on several neutral current four-lepton interactions involving muons providing a novel way for the indirect search of New Physics. Our results, summarized in Table~\ref{tab:operators}, constrain effective Wilson coefficients such as $\left| [c_{LL}]_{\mu\mu\alpha\alpha}\right|/v^2 \allowbreak \leq 1.0~\cdot~10^{-2}~\text{GeV}^{-2}$, for the ongoing  NA64$\mu$ analysis, with projections extending to $\left| [c_{LL}]_{\mu\mu\mu\mu}\right|/v^2 \leq \allowbreak  1.7 \cdot 10^{-4}~\text{GeV}^{-2}$ or $\left| [c_{LL}]_{\mu\mu\tau\tau}\right|/v^2 \leq  1.3 \cdot 10^{-4}~\text{GeV}^{-2}$, at $1.0\cdot10^{14}$~MOT, corresponding to New Physics scales up to 
$\Lambda \sim 90~\text{GeV}$. In the case of $\nu$SMEFT the constraint on the Wilson coefficient can reach up to $\left|C_{LN}^\mu\right|/\Lambda^2 \leq  4.8\cdot 10^{-3}~\text{GeV}^{-2}$ for the current accumulated MOT, and up to $\left|C_{LN}^\mu\right|/\Lambda^2 \leq  1.0\cdot 10^{-4}~\text{GeV}^{-2}$ or $\Lambda \gtrsim 100$~GeV, for the planned upgrade of NA64$_\mu$. 
As demonstrated in this work, NA64$\mu$ offers the possibility of directly exploring four-fermion interactions involving muons, including previously unconstrained flavor-violating SMEFT operators with $\alpha \neq \beta$, for which we obtain a comparable sensitivity. Furthermore, our SMEFT results can be straightforwardly reinterpreted in terms of a WEFT associated with new BSM states whose masses lie between the GeV and electroweak scales. These direct constraints therefore provide an important complement to those expected from possible future muon collider experiments~\cite{InternationalMuonCollider:2024jyv}. Unlike NA64$\mu$, which probes low-energy scattering processes sensitive to non-resonant New Physics via contact operators, muon colliders such as the proposed multi-TeV machines will have the capability to produce heavy mediators on shell~\cite{Aime:2022flm}.

\section*{Acknowledgments}
We would like to thank Dr. Henri Sieber for very useful discussions at the early stage of this work. We would also like to thank Javier Fuentes-Martín and Martín González-Alonso for useful discussions. This work has been partially supported by the Spanish Severo
Ochoa Excellence grant CEX2023-001292-S
(AEI/10.13039/501100011033). \\VML  acknowledges financial support by Prometeo CIPROM/2021/054 from Generalitat Valenciana  and by the Spanish grants PID2023-147306NB-I00. \\ JLP and JHG acknowledge financial support from European Union’s Horizon 2020 research and innovation program under the Marie Skłodowska-Curie grants HORIZON-MSCA-2021-SE-01/101086085-ASYMMETRY and H2020-MSCA-ITN-2019/860881-HIDDeN, from the Spanish Research Agency (Agencia Estatal de Investigaci\'on) through grants PID2023-148162NB-C21 and PID2022-137268NA-C55, and CNS2022-136013 funded by \\MICIU/AEI/10.13039/501100011033 and by “European Union NextGenerationEU/PRTR'', and from the MICIU with funding from the European Union NextGenerationEU (PRTR-C17.I01) and Generalitat Valenciana (ASFAE/2022/020). LMB acknowledges the support from the Spanish Research Agency (Agencia Estatal de Investigaci\'on) through grants RyC-030551-I, PID2021-123955NA-100 and CNS2022-135850 funded by MCIN/AEI/FEDER, UE (Spain). 

\appendix

\section{Production cross section}
\label{app:XS}


\subsubsection{$\mathcal{O}_{LL}$ and $\mathcal{O}_{L\ell}$}
For the effective SMEFT operators we are going to compute the amplitudes in terms of the $\varepsilon_{\alpha\beta}^{\mu,V}$ and $\varepsilon_{\alpha\beta}^{\mu. A}$ parameters. The mapping to the SMEFT parameters is given in Eq.~(\ref{eq:weft}). The squared amplitude summed and averaged over spins can be written in terms of $\mathcal{C}_n$ coefficients as in Eq.~\eqref{eq:dsdxdk2}. The differential cross section written in terms of these coefficients is:
\begin{eqnarray}
        \left.\frac{d\sigma}{dxdk^2}\right|_{\mathcal{O}_{LL,L\ell}} &=& \frac{\alpha^2}{48\pi^3v^4}\frac{1-x}{x}\sqrt{x^2-\frac{k^2}{E_\mu^2}}\int \frac{d\tilde{u}}{\tilde{u}^2}\,\,\chi^{\rm WW}\times \nonumber \\
&\times&\sum_{\alpha,\beta}\left[\left|\delta_{\alpha\beta}\left(\delta_{\alpha\mu}-\frac{1}{2}+2s_{\rm w}^2\right)+\varepsilon^{\mu,V}_{\alpha\beta}\right|^2\sum_{n=1}^3\frac{\mathcal{C}_n^{V}}{\tilde{u}^{n-1}} \right.+ \\
&+&\left.\left|\delta_{\alpha\beta}\left(\delta_{\alpha\mu}-\frac{1}{2}\right)+\varepsilon^{\mu,A}_{\alpha\beta}\right|^2\sum_{n=1}^3\frac{\mathcal{C}_n^{A}}{\tilde{u}^{n-1}}\right] \nonumber
       \label{eq:sigmaSMEFT}
\end{eqnarray}
with 
\[
\begin{array}{rlrl}
\mathcal{C}_1^{V} &= k^2\dfrac{(x-2)x+2}{1-x}\,, 
& \quad \mathcal{C}_1^{A} &= k^2\dfrac{2m_{\mu}^2x^2+(x-2)x+2}{1-x}\,, \\[1ex]
\mathcal{C}_2^{V} &= 2k^2x(k^2+2m_\mu^2)\,, 
& \quad \mathcal{C}_2^{A} &= 2k^2x(k^2-4m_\mu^2)\,, \\[1ex]
\mathcal{C}_3^{V} &= 2k^2(k^2+2m_\mu^2)(k^2(1-x)+m_\mu^2x^2)\,, 
& \quad \mathcal{C}_3^{A} &= 2k^2(k^2-4m_\mu^2)(k^2(1-x)+m_\mu^2x^2)\,.
\end{array}
\]

The SM contribution to the process is given by the terms proportional to $\delta_{\alpha\beta}$ inside the brackets. Note that there is an interference term between the SM and SMEFT contributions. However, its impact is not very significant since NA64$\mu$ probes scales around the electroweak scale.

\subsubsection{$\mathcal{O}_{LN}^{\alpha}$ and $\mathcal{O}_{ff}$}

The $\mathcal{O}_{LN}^{\alpha}$ and $\mathcal{O}_{ff}$ operators lead to the same expression for the differential cross section given by

\begin{eqnarray}
        \left.\frac{d\sigma}{dxdk^2}\right|_{\mathcal{O}_{LN,ff}}&=& \left.\frac{d\sigma}{dxdk^2}\right|_{\mathcal{O}_{LL,L\ell}(\varepsilon=0)}+ \\
        &+&\frac{\alpha^2}{48\pi^3}\left|\frac{C_i}{\Lambda^2}\right|^2\frac{1-x}{x}\sqrt{x^2-\frac{k^2}{E_\mu^2}}\sqrt{1-\frac{4m_N^2}{k^2}}\int \frac{d\tilde{u}}{\tilde{u}^2}\,\,\chi^{\rm WW}\sum_{n=1}^3\frac{\mathcal{C}_n^{LN,ff}}{\tilde{u}^{n-1}},\nonumber
 \label{eq:sigmanuSMEFTnc}
 \end{eqnarray}
where the first term is the SM contribution and the coefficient $C_i$ is given by $C_{\mu\mu}$ ($C_{LN}^\mu$) for $\mathcal{O}_{LN}^\alpha$ ($\mathcal{O}_{ff}$). The coefficients $\mathcal{C}_n^{LN,ff}$ 
are given by:
\begin{eqnarray}
    \mathcal{C}_1^{LN,ff}&=&k^2\left(1+\frac{2m_N^2}{k^2}\right)\left[\frac{(x-2)x+2}{1-x} + \frac{m_\mu^2x^2}{k^2(1-x)}\right]\,,\\
    \mathcal{C}_2^{LN,ff}&=&k^2\left(1+\frac{2m_N^2}{k^2}\right)2x(k^2-m_\mu^2)\,, \\
    \mathcal{C}_3^{LN,ff}&=&k^2\left(1+\frac{2m_N^2}{k^2}\right)2(k^2-m_\mu^2)(k^2(1-x)+m_\mu^2x^2)\,,
\end{eqnarray}
and where we have assumed that $N$ is a Dirac field.

\subsubsection{$\mathcal{O}^{\alpha\beta}_{LNL\ell}$}
Finally we can write the expression of the differential cross section for the scalar operator as:
\begin{eqnarray}
        \left.\frac{d\sigma}{dxdk^2}\right|_{\mathcal{O}_{LNL\ell}}&=& \left.\frac{d\sigma}{dxdk^2}\right|_{\mathcal{O}_{LL,L\ell}(\varepsilon=0)}+\\
        &+&\frac{\alpha^2}{48\pi^3}\left|\frac{C_{LNL\mu}^{\alpha\beta}}{\Lambda^2}\right|^2 \frac{1-x}{x}\sqrt{x^2-\frac{k^2}{E_\mu^2}}\sqrt{1-\frac{m_N^2}{k^2}}\int \frac{d\tilde{u}}{\tilde{u}^2}\,\,\chi^{\rm WW}\sum_{n=1}^3\frac{\mathcal{C}_n^{LNL\ell}}{\tilde{u}^{n-1}},\,\nonumber
        \label{eq:sigmanuSMEFTcc}
\end{eqnarray}
where the first term is the SM contribution and the coefficients $\mathcal{C}_n^{LNL\ell}$ are given by:
\begin{eqnarray}
    \mathcal{C}_1^{LNL\ell}&=&\frac{3}{4}k^2\left(1-\frac{m_N^2}{k^2}\right)\frac{x^2}{2(1-x)}\,,\\
    \mathcal{C}_2^{LNL\ell}&=&\frac{3}{4}k^2\left(1-\frac{m_N^2}{k^2}\right)x(k^2-2m_\mu^2)\,,\\
    \mathcal{C}_3^{LNL\ell}&=&\frac{3}{4}k^2\left(1-\frac{m_N^2}{k^2}\right)(k^2-2m_\mu^2)(k^2(1-x)+m_\mu^2x^2)\,. 
\end{eqnarray}



%

\end{document}